\documentclass[12pt,fleqn]{iopart}
\expandafter\let\csname equation*\endcsname\relax 
\expandafter\let\csname endequation*\endcsname\relax 

\usepackage{geometry}
\usepackage{graphicx}
\usepackage{amsmath}
\usepackage{amssymb}
\usepackage{physics}
\usepackage{mathrsfs}
\usepackage{hyperref}

\usepackage[charter]{mathdesign}
\DeclareMathAlphabet{\pazocal}{OMS}{zplm}{m}{n}
\def\cal#1{\pazocal{#1}} 
\def\d{{\rm d}}
\def\i{{\rm i}}
\def\e#1{{\rm e}^{#1}}
\def\del#1{\delta\hspace{-0.08em}#1}

\begin{document}

\title{Non-adiabatic ionization with tailored laser pulses}
\author{Sajad Azizi, Ulf Saalmann, Jan M Rost}
\address{Max-Planck-Institut f{\"u}r Physik komplexer Systeme\\
 N{\"o}thnitzer Str.\ 38, 01187 Dresden, Germany}
\date{\today}

\begin{abstract}\noindent
Non-adiabatic photo-ionization is difficult to control as it relies on the derivatives of the envelope and not on phase-details of the short ionizing pulse. 
Here, we introduce a catalyzing state, whose presence render non-adiabatic ionization sensitive to phase-details of tailored pulses. 
Since a catalyzing state is in general easy to create, this opens a perspective for coherent control of ultra-fast ionization.
\end{abstract}

\submitto{\jpb}
{\def\newpage{\relax}\maketitle}

\section{Introduction}
\noindent
In recent years new phenomena in coupling of light to matter have been uncovered through tailored laser fields, where the emphasis has shifted from a typical coherent-control scenario by a shaped laser pulse \cite{shbr12} to two-color pulses and/or different time-dependent polarizations \cite{eieg+95,eh01,lewo17}.
Coherent control of multi-photon transitions in the optical strong-field regime by shaped pulses has been demonstrated \cite{mesi98mesi99} aided by Stark shifts which modify multi-photon processes \cite{stdu+93}.
Along another thrust, ever shorter pulses with nominal carrier frequencies in the extreme-ultraviolet (XUV) regime have been pursued, either generated by high-harmonic sources \cite{lire+17,gaja+17} or by free-electron lasers \cite{haha+18,sege+18}, which can produce quite intense pulses. 
For those pulses, phase manipulation is also possible \cite{wibr+20}. 
Surprisingly, using the longitudinal coherence within the waveform of light wave-packets, produced by individual relativistic electrons, it is even possible with synchrotrons to shape pulses on the attosecond time-scale (duration and separation) with XUV carrier frequencies \cite{hika+19}.

For the regime of ultra-short intense pulses, we have demonstrated non-adiabatic photo-ionization (NAPI) \cite{toto+09,tosa+15,nisa+18}, typically for weakly-bound systems $E_0\,{\ll}\,\omega$, with the electron's binding energy $E_0$ and the photon frequency $\omega$. The characteristic of NAPI is a peak of the ionization yield just above the ionization threshold. The physics behind NAPI is a time-scale hierarchy such that the photo-electron cannot follow the fast change of the pulse envelope (therefore non-adiabatic photo-ionization).
We have shown that NAPI is sensitive to the derivative of the pulse envelope \cite{nisa+18}. As a consequence, a single Gaussian pulse acts like a double pulse in the NAPI regime with a time delay between the two pulse-derivative peaks given roughly by the width of the original Gaussian pulse.
 
In the following, we will investigate if NAPI can be influenced and steered by tailored pulse forms, in analogy to coherent control of standard (single) photo-ionization.
To be specific, yet paradigmatic, we use a pulse form with three control parameters $a$, $\tau$ and $\phi$, that is routinely used in coherent control experiments \cite{mesi98mesi99,wopr+06,bawo+08}. 
It is generated in the frequency domain by a modulation of the spectral phase in the vector potential
\begin{equation}\label{eq:train-omega}
 A(\omega')=N_{T}\frac{F}{\omega}\e{-[\omega'-\omega]^2T^2/8\ln2}\e{\i\, a\,{\sin}(\omega'\tau+\phi)}
\end{equation}
of a pulse with peak field strength $F$, carrier frequency $\omega$ and full-width-at-half-maximum duration $T$.
Hereby $N_{T}\,{\equiv}\,T/2\sqrt{2\pi\ln2}$ ensures the proper amplitude of the corresponding pulse.

\begin{figure}[t!]
 \centering
 \includegraphics[width=0.8\columnwidth]{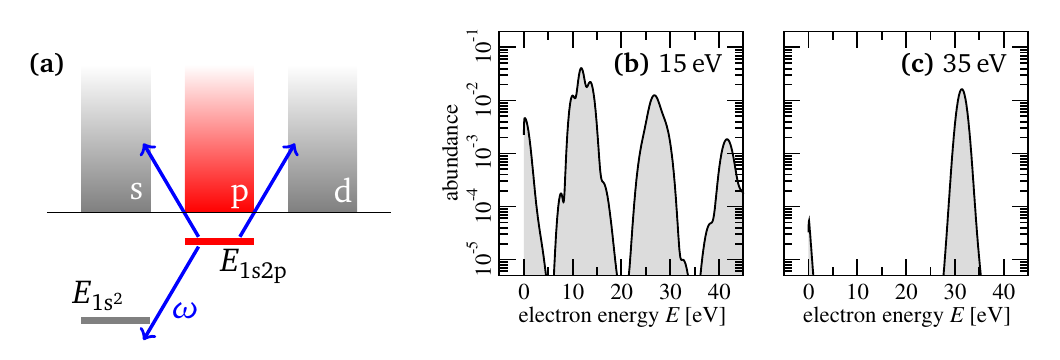}
 \caption{{\bf (a)} Sketch of the physical system with the relevant states. 
 {\bf (b,\,c)} Photo-electron spectra in logarithmic scale from short intense pulses ($I{=}10^{16}$W/cm$^2$, $T{=}1$\,fs) at two photon frequencies $\omega$.}
 \label{fig:sketch}
\end{figure}

We do not aim at a specific control target, e.\,g., maximizing or minimizing the population of a specific state. 
Rather, we want to identify situations where the NAPI spectrum depends sensitively on pulse details, in particular the modulation phase $\phi$. We will see that this requires another discrete state to be closely coupled, which acts as a ``catalyzer'' to evoke controllability of NAPI.
To this end, we study ionization from the excited 1s2p state of Helium.
Changing the photon frequency $\omega$, non-resonant as well as resonant situations are realized by coupling to a deeper-lying bound state as indicated by the sketch in Fig.\,\ref{fig:sketch}a.
A resonant coupling can strongly enhance the Stark shift and thereby drive non-adiabatic ionization. We determine the electron dynamics in a single active-electron description as detailed in \ref{sec:num}.

\section{Ionization by single Gaussian pulses}\label{sec:gauss}
To set the stage and put NAPI into perspective, we show the photo-electron spectra for single Gaussian pulses
\begin{equation}\label{eq:gauss}
 A(t) = \frac{F}{\omega} \e{-2\ln2 \ t^2/T^2}\cos(\omega t), 
\end{equation}
for two different photon frequencies $\omega$ in Fig.\,\ref{fig:sketch}b,c. 
For the smaller one ($\omega\,{=}\,15$\,eV) one can distinguish four peaks corresponding to the absorption of $j=0\ldots3$ photons within the energy range shown. 
For future reference, we define energy intervals $\Delta E_j$ about these peaks with $\Delta E_j=\{E\,|\,{-}\omega/2{<} E{-}E_j{<}\omega/2 \} $ with $E_j\equiv E_0+j\omega$ reached by $j$ photons from the initial state at $E_{0}$. Note, 
that the final electron states can carry different angular momentum $\ell$ in this few-photon scenario, see the sketch in Fig.\,\ref{fig:sketch}a.

\begin{figure}[t!]
 \centering
 \includegraphics[width=0.7\columnwidth]{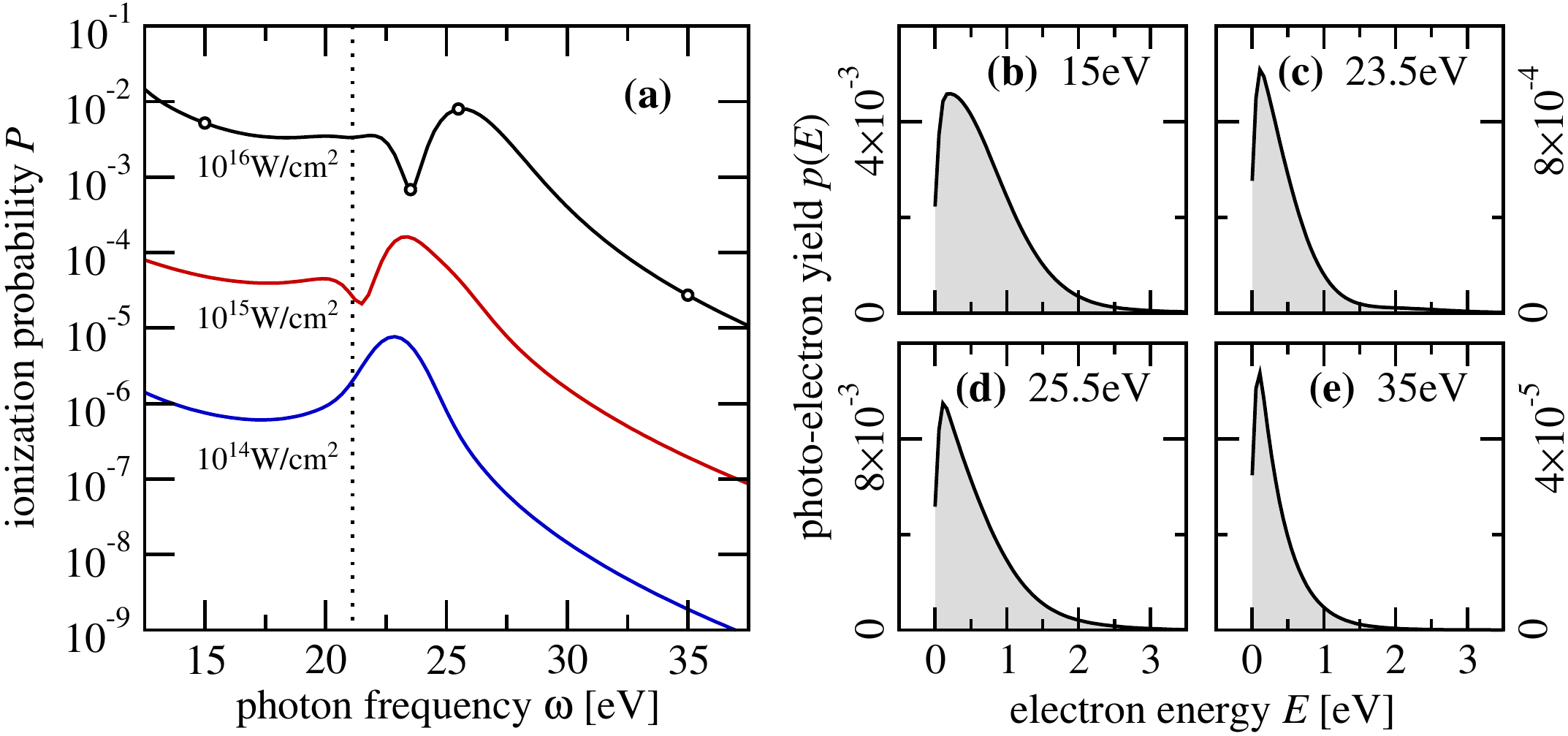}
 \caption{{\bf (a)} Ionization probability for the non-adiabatic channel as a function of the photon frequency $\omega$ for a Gaussian pulse with duration $T{=}1$\,fs and various intensities specified in the graph. The dotted line marks the transition energy between the ${\rm 1s^{2}}$ and $\rm 1s2p$ state, cf.\ Fig.\,\ref{fig:sketch}. 
 {\bf (b--e)} Photo-electron spectra for selected frequencies specified in the graphs and marked in panel (a) with circles.}
 \label{fig:gauss}
\end{figure}%
The NAPI channels correspond to ``zero-photon absorption'' represented by low-energy electrons in the interval $\Delta E_0$, with $E\,{>}\,0$, forming the first peak in the spectra of Fig.\,\ref{fig:sketch}b. The next peak around $E_1$ is the biggest one and corresponds to single-photon ionization into $\ell\,{=}\,0,2$ continua. It is structured through dynamic interference as it is well known \cite{basa+17}. Also clearly visible on the logarithmic scale are the peaks in the intervals $\Delta E_2$ and $\Delta E_3$, respectively.
At the higher photon energy of $\omega\,{=}\,35$\,eV, the light-matter interaction is basically perturbative such that only the (standard) single-photon ionization peak $E_1$ survives, even on the logarithmic scale.

With Fig.\,\ref{fig:gauss} we explore the total NAPI probability $P$, as defined in Eq.\,\eqref{eq:prob}, for continuum electrons with p-character (angular momentum $\ell{=}1$) for an ultra-short ($T{=}1$\,fs) Gaussian pulse as a function of $\omega$ for three different laser intensities. The p-state probabilities dominate since the NAPI process
is an effective zero-photon process with a (small) admixture of an even number of photons.
Hence, optical selection rules do not permit $\ell{=}0,\,2$ final states to be reached from our initial p-state, and the allowed final f-state channel ($\ell{=}3$) is much weaker than the final p-state channel.

Outside resonances, photo-ionization yields typically decrease with increasing frequency, for large $\omega$ proportional to $\omega^{-7/2}$ \cite{ro95}, which is also the case here. 
However, in the frequency range displayed, the spectrum is dominated by a resonance-like peak between 20 and 25\,eV. 
For increasing intensity, it shifts slightly to larger $\omega$ and develops a preceding dip. 
This structure is due to the resonance with the 1s$^2$--1s2p electron transition located (for weak fields) at $E_{\rm 1s2p}-E_{\rm 1s^{2}}\approx21.1$\,eV.
Despite the strong variation of the yield around frequencies close to the resonance,
the corresponding NAPI spectra (Fig.\,\ref{fig:gauss}b--e) have remarkably similar and structure-less shapes inside and outside the resonance region, albeit on very different scales.

\begin{figure}[t!]
 \centering
 \includegraphics[width=0.7\columnwidth]{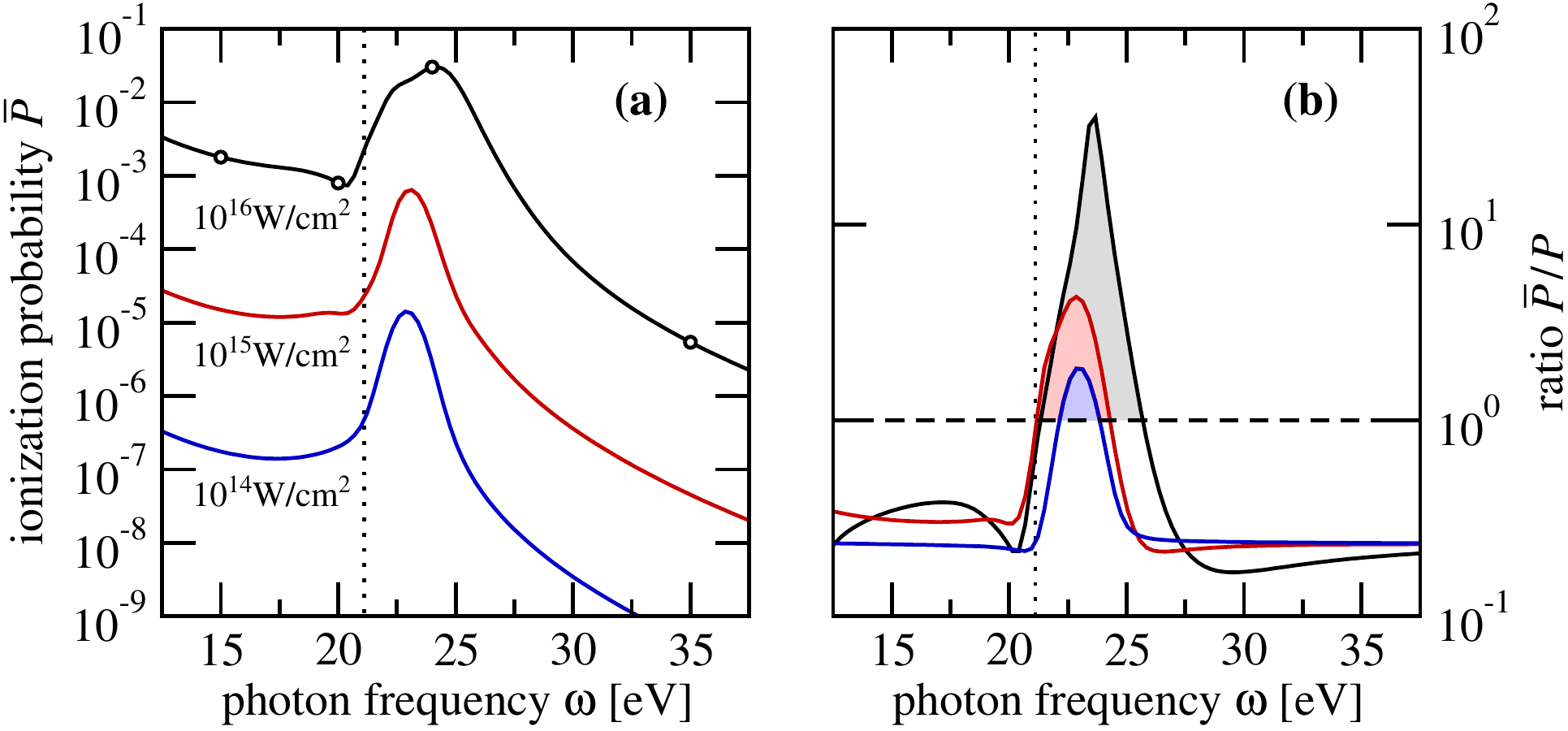}
 \caption{{\bf (a)} Same as Fig.\,\ref{fig:gauss}a but for a pulse train, given by Eq.\,\eqref{eq:train}, with $a{=}1$, $\tau{=}4$\,fs.
 The curves are averages over the phase parameter $\phi$. 
 {\bf (b)} Ratios of the probabilities from panel (a) and Fig.\,\ref{fig:gauss}a as a function of the photon frequency $\omega$.
 The shadings mark the enhancement region.}
 \label{fig:ratio}
\end{figure}%
\section{Sensitivity of non-adiabatic photo-ionization to the modulation phase}\label{sec:train}
One might infer from the quite similar shapes of the NAPI spectra in Figs.\,\ref{fig:gauss}b--e that it is very difficult to coherently control NAPI with standard shaped pulses.
Yet, as it will turn out, a pulse train with a modulated spectral phase,
routinely used in experimental realizations of pulse shaping in the frequency domain 
\cite{mesi98mesi99,wopr+06,bawo+08} can achieve controllability of NAPI. 
The pulse train is obtained from a Fourier transform $A(t)={\cal F}[A(+\omega')+A^{*}(-\omega')]/2$ with $A(\omega')$ given in Eq.\,\eqref{eq:train-omega}
and reads \cite{woas+12}
\begin{equation}\label{eq:train}
 A(t)=\frac{F}{\omega}\sum_{k}J_{k}(a)\;\e{-2\ln2[t-k\tau]^2/T^2}{\cos}\,(\omega[t{-}k\tau]-k\phi),
\end{equation}
with $J_k$ denoting Bessel functions.
How strongly the original Gaussian pulse is distributed over separate pulses in the train is controlled by $a$, the amplitude of the phase oscillation in Eq.\,\eqref{eq:train-omega}.
We will choose $a{=}1$, which results in a train with essentially 9 pulses, since 
$J_{0\ldots4}(1)\,{\approx}\,\{0.765,0.440,0.115,0.020,0.002\}$.
The delay between the pulses is fixed by $\tau$.
The modulation phase $\phi$ introduces a difference $\phi$ in the carrier-envelope phase of adjacent pulse members of the train, see Eq.\,\eqref{eq:train}, which will become important later on.

Firstly, we take a look at the ionization probability with this pulse train as we did in Fig.\,\ref{fig:gauss}a for single Gaussian pulses.
Figure~\ref{fig:ratio}a shows the probability $\overline{P}$ as a function of $\omega$ for the same three intensities $I$. Note that the separation $\tau\,{=}\,4$\,fs together with the duration $T\,{=}\,1$\,fs of the individual pulses in the train ensures that they do not overlap in time.
The bar indicates that we have averaged the spectra over the modulation phase $\phi$, 
\begin{equation}\label{eq:avespec}
 \overline{p}(E)=\frac{1}{2\pi} \int\!\mathrm{d}\phi\, p_{\phi}(E)\,,
\end{equation}
with $\overline{P}$ being the integral over $\overline{p}(E)$, cf.\ Eq.\,\eqref{eq:prob}.
Despite the different shapes compared to the single-pulse yields in Fig.\,\ref{fig:gauss}a, the qualitative behavior is the same: a monotonic decrease interrupted by a peak in the vicinity of $\omega\approx E_{\rm 1s2p}{-}\,E_{\rm 1s^{2}}$.
Interestingly, the total yield can be \emph{considerably} larger than for the Gaussian pulse, as apparent from Fig.\,\ref{fig:ratio}b, which shows the ratio of the yields from Fig.\,\ref{fig:ratio}a and Fig.\,\ref{fig:gauss}a.
Since NAPI is enhanced by large derivatives of the pulse envelope \cite{nisa+18}, it is surprising that a longer pulse, with smaller slopes in the overall envelope, can induce an order-of-magnitude larger ionization probability at $I\,{=}\,10^{16}$W/cm$^2$. The enhancement for all three intensities is visualized by shaded areas in Fig.\,\ref{fig:ratio}b. 

\begin{figure}[t!]
 \centering
 \includegraphics[width=0.7\columnwidth]{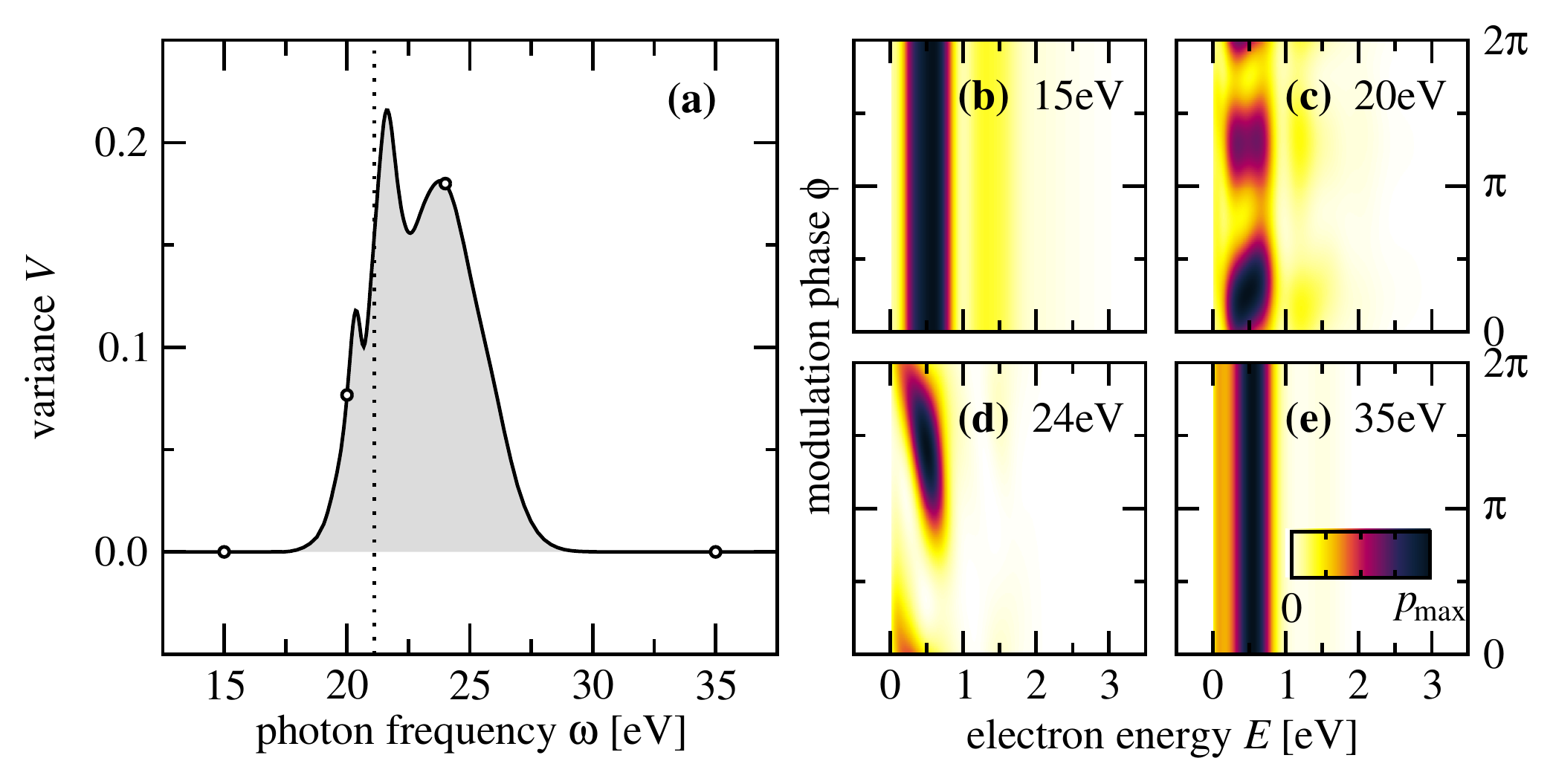}
 \caption{{\bf (a)} Variance of the photo-electron spectra with respect to the modulation phase $\phi$, defined in Eq.\,\eqref{eq:vari}, for pulse trains according to Eq.\,\eqref{eq:train} with $I{=}10^{16}$W/cm$^{2}$, $a{=}1$, $\tau{=}4$\,fs as a function of the photon frequency $\omega$. {\bf (b--e)} Representative photo-electron spectra $p_{\phi}(E)$ as a function of $\phi$ for selected photon frequencies, marked with circles in Figs.\,\ref{fig:ratio}a and \ref{fig:variance}a, with the color scale for the electron yield normalized to the maximal yield.}
 \label{fig:variance}
\end{figure}%
Secondly, we want to assess how strongly the photo-electron spectra depend on the phase parameter $\phi$, since a strong sensitivity could represent a knob for controlling NAPI.
In order to quantify this sensitivity, we compute the variance of the spectra with respect to the modulation phase $\phi$, defined as
\begin{equation}\label{eq:vari}
V^{2} = \frac{1}{\overline{P}{}^{2}}\int\!\mathrm{d}E\, \int\!\mathrm{d}\phi\, \big[p_{\phi}(E) - \overline{p}(E)\big]^{2}
\end{equation}
with $\overline{p}(E)$ from \eqref{eq:avespec} above, and show it in Fig.\,\ref{fig:variance}a.
As already seen for the probability, the region around the resonance sticks out and shows a noticeable variance.
Below and above the resonance frequency, however, one sees the behavior characteristic for NAPI, namely that the ionization is determined by the envelope of the pulse only \cite{nisa+18}, but not by any carrier-oscillation features.
In order to visualize this dependence, we show representative photo-electron spectra from different regions appearing in Fig.\,\ref{fig:ratio}a in parallel to Fig.\,\ref{fig:variance}b--e now, however, as a function of electron excess energy and modulation phase $\phi$.
While for a single Gaussian pulse, the photo-electron spectra are not affected by the resonance, the situation is quite different for the spectra generated with the pulse train: Here, the shape of the spectra varies strongly around the resonance for different $\phi$.

\section{The role of the catalyzing state}\label{sec:role}
\begin{figure}[t!]
 \centering
 \includegraphics[width=0.8\columnwidth]{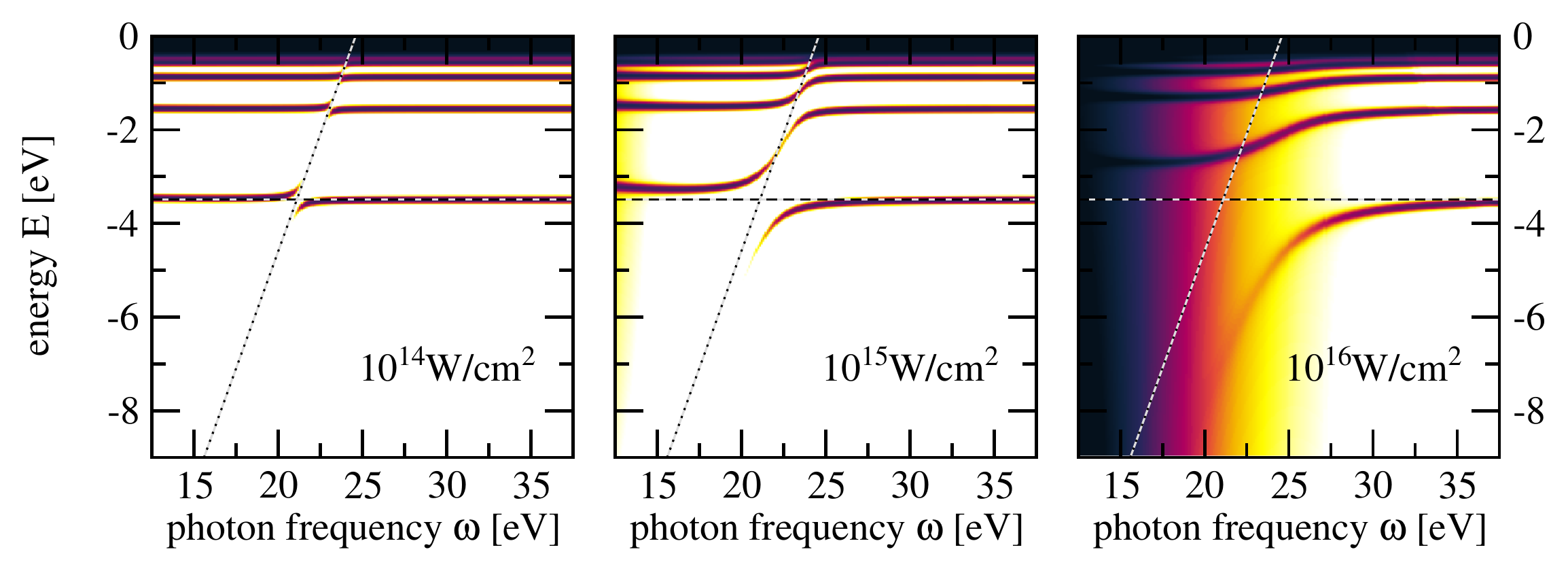}
 \caption{Density of states with p-character ($\ell{=}1$) in logarithmic scale from a dressed-state dia\-go\-na\-li\-zation, cf.\ \ref{sec:dress}, as a function of the photon frequency $\omega$ for three laser intensities.
 The dashed horizontal line marks the energy $E_{\rm 1s2p}$, the dotted sloping line the energy $E_{\rm 1s^{2}}{+}\,\omega$.
 They cross at resonance.}
 \label{fig:dos}
\end{figure}

Having established that NAPI can be coherently controlled, at least in the presence of a catalyzing state which can be energetically very far away (here at an energetic distance of $\omega$), we will elucidate the origin of the sensitivity of NAPI on $\phi$ in the vicinity
of the resonance. To this end, we show in Fig.\,\ref{fig:dos} how energies of p-state ($\ell{=}1$) electrons get ``deformed'', i.\,e.\ Stark shifted and hybridized, due to the coupling to s- and d-states. 
The color code of the hybridized density of states (DOS) marks the strength of their p-character at energy $E$ (for a specific $\omega$), for details see \ref{sec:dress}. 
The color-coded DOS also nicely illustrates the hybridization of angular momentum character of the DOS near the avoided crossings:
Along an adiabatic trace which bends strongly near the avoided crossing, the character of the electron density changes, from dominant s-character through the 
$1{\rm s}^2+\omega$ dressed state (with a finite slope due to $\omega$) to the 1s2p state with dominant p-character given by a horizontal line at energy $E_{\rm 1s2p}$ which the electron density trace approaches towards large frequencies from below (and towards small frequencies from above).
Note that the background density is also hybridized, most clearly visible from the area-filling color shades in Fig.\,\ref{fig:dos}.

At resonance $E_{\rm 1s2p}\,{=}\,E_{\rm 1s^2}\,{+}\,\omega$ the 1s2p-state shows an Autler-Townes splitting \cite{auto55}, i.\,e., structures below and above the field-free energy of $E_{\rm 1s2p}\,{=}\,{-}3.48$\,eV (dashed line).
The latter is crossed by the dressed state with field-free energy $E_{\rm 1s^2}+\omega$ (also shown with a dashed line). 
Note that the actual field-dressed states are shifted and have their interaction-caused avoided crossing at higher photon energies than the field-free states. 
This results in peaks consistently blue shifted with respect to the $\omega=E_{\rm 1s2p}-E_{\rm 1s^2}$ resonance energy of 21.1\,eV in Figs.\,\ref{fig:gauss}a\,--\,\ref{fig:variance}a. 
This blue shift is another signature of non-adiabatic ionization, which is in fact a virtual two-photon process:
Whereas the coupling from the initial state to the catalyzing state (``1st photon'') is symmetric around the resonance condition, the transition from the catalyzing state to the continuum (``2nd photon'') is not. 
This can be illustrated in 2nd-order perturbation theory, neglecting for convenience the dipole-coupling matrix elements. The transition probability to a continuum state at energy $E$ reads $p_\omega(E)\propto\exp(-[[\delta_1{-}\omega]^{2}+[\delta_2{-}\omega]^{2}]T^2/4\ln2)$, with the transition energies $\delta_1\equiv E_{\rm 1s2p}{-}\,E_{\rm 1s^2}$ and $\delta_2\equiv E\,{-}\,E_{\rm 1s^2}$, respectively.
This probability does not peak at $\omega\,{=}\,\delta_1$ but rather at $\omega\,{=}\,[\delta_1{+}\delta_2]/2$. Noting that the non-adiabatic photo-electrons have an energy around $E\,{=}\,1$\,eV, cf.\ Figs.\,\ref{fig:gauss}b--e, gives for our system an optimal frequency of about $\omega\,{\approx}\,23.4$\,eV, in accordance with the numerical results presented in the previous sections.

\section{Summary}
We have investigated how non-adiabatic photo-ionization (NAPI) induced by ultra-short XUV pulses, can be influenced through specific pulse forms to exert coherent control as well-known for standard photo-ionization. 
We have demonstrated that a spectral-phase modulated pulse train, where individual pulses $k$ in the train have different carrier-envelope phases $k\phi$, can achieve control \emph{provided} an additional catalyzing state is available. 
A state qualifies as catalyzing if it forms a resonance with the initial state $E_0-E_{\rm cat}\,{\approx}\,\omega$. 
Since this kind of pulse trains is routinely used to shape longer optical pulses in the frequency domain and since pulse shaping has also been demonstrated for XUV pulses recently, we expect that controlling NAPI will be possible experimentally in the future. 

Moreover, while illustrated here with the quantitative example of Helium, the control scheme should be applicable to other targets as well, in particular as it relies on a resonant situation which usually dominates over other (e.\,g.\ multi-electron effects) in the vicinity of the resonant energy.

\section*{Acknowledgements}
This work has been supported by the Deutsche Forschungsgemeinschaft (DFG) through the priority program 1840 ``Quantum dynamics in tailored intense fields''.

\appendix
\section{Numerical details}\label{sec:num}
The Helium atom is treated in the single-active-electron approximation.
The following effective potential \cite{zi15}
\begin{align}
V(r) = -\frac{1+\mathrm{e}^{-\alpha r}}{r}
\end{align}
with $\alpha=2.1325$ has been used.
It provides a good approximation for the energies of the relevant states of Helium, namely $E_{\rm 1s^{2}}{=}-24.59$\,eV and $E_{\rm 1s2p}{=}-3.48$\,eV (with the values $E_{\rm 1s^{2}}{=}-24.59$\,eV and $E_{\rm 1s2p}{=} -3.37$\,eV from accurate two-electron calculations \cite{sc98}).

The lowest $j_{\rm max}$ field-free states $\varphi_{j\ell}$ of each angular momentum $\ell = 0 \ldots\ell_{\rm max}$ are calculated numerically by means of the Numerov method in a finite box $r = 0 \ldots r_{\rm max}$ with a grid spacing of $\del{r} = 0.01$\,a$_0$, which gives highly accurate energies and states. The box size $r_{\rm max} = 2000$\,a$_0$ is chosen such that the continuum wave-packet does not reach the box boundary within the propagation time. We use $\ell_{\rm max}{=}\,4$ and $j_{\rm max}{=}\,1500$, i.\,e.\ 7500 states in total, with the highest ones having an energies of $E\,{\approx}\,75$\,eV. 

The time-dependent Schr\"odinger equation (TDSE) is solved in velocity form for a linearly-polarized pulse in terms of the field-free basis
\begin{equation}
 \i\dot{a}_{j\ell}(t)=\sum_{j'\ell'}\big[E_{j\ell}\delta_{jj'}\delta_{\ell\ell'}+A(t)D_{j\ell j'\ell'}\big]a_{j'\ell'}(t)
\end{equation}
with the coupling matrix $D_{j\ell j'\ell'}\equiv\langle\varphi_{j\ell}|\widehat{p}_{z}|\varphi_{j'\ell'}\rangle$ and the time-dependent vector potential $A(t)$ chosen appropriately, cf.\ Eqs.\,\eqref{eq:gauss} and \eqref{eq:train}.
As a convergence check we have solved the TDSE for some selected cases in the length form, whereby $A(t)\to F(t)=-\frac{\d}{\d t}A(t)$
and $D_{j\ell j'\ell'}\to\langle\varphi_{j\ell}|z|\varphi_{j'\ell'}\rangle$, with essentially identical results.

The electron energy spectrum for a certain angular momentum $\ell$ (here only $\ell{=}1$, apart from Fig.\,\ref{fig:sketch}) is obtained with $w_{j}\equiv|a_{j,\ell=1}(t\mbox{${\to}\infty$})|^2$ as 
\begin{equation}\label{eq:spec}
 p(E)=\frac{1}{\sqrt{\pi}\,\del{E}} \sum_{j}^{(\ell=1)} w_{j}\e{-[E-E_{j\ell_{j}}]^2/\del{E}^2},
\end{equation}
with $\del{E}{=}0.05$\,eV.
The corresponding non-adiabatic ionization probability are obtained from the integration
\begin{equation}\label{eq:prob}
 P=\int_{0}^{E_{1/2}} \!\!\!\!\d E\, p(E)
\end{equation}
with the upper integration limit $E_{1/2}=E_{0}+\omega/2$.

\section{Dressed-state description}\label{sec:dress}
In order to analyze the role of the catalyzing state, cf.\ Sect.\,\ref{sec:role},
we build a dressed-state matrix \cite{sagi+18}
\begin{equation}\label{eq:dsham}
H_{jj'}(F,\omega)=[E_{j}+n_{j}\omega]\delta_{jj'}+\frac{F/\omega}{2}D_{j\ell_{j} j'\ell_{j'}},
\end{equation}
that has a block-diagonal shape, whereby the 5 blocks are defined by the ``photon numbers'' $n_{j}=\{+1,+1,0,-1,-1\}$ and the angular momenta $\ell_{j}=\{0,2,1,0,2\}$.
Field strength $F$ and photon energy $\omega$ are parameters here. 
Field-free eigenenergies $E_j$ and the coupling-matrix elements $D_{j\ell_{j} j'\ell_{j'}}$ are calculated as described in \ref{sec:num}.

The eigenstates from the diagonalization of matrix \eqref{eq:dsham}
\begin{equation}\label{eq:dseig}
\sum_{j'}H_{jj'}(F,\omega)V_{j'k}(F,\omega)=V_{jk}(F,\omega)E_{k}(F,\omega)
\end{equation}
are used to calculate by means of a angular-momentum projection operator $P^{[\ell]}_{jj'}\equiv\delta_{\ell\ell_j}\delta_{jj'}$ weights $w_{k}=\sum_{jj'}V_{jk}P^{[1]}_{jj'}V_{j'k}$ and therewith, according to Eq.\,\eqref{eq:spec}, the density of states (DOS) shown in Fig.\,\ref{fig:dos}.

\clearpage
\section*{References}
\providecommand{\newblock}{}

\end{document}